\documentclass[12pt]{article}
\usepackage[body={18cm, 23cm, centered}]{geometry}
\usepackage{amsmath,amssymb,amsthm,amscd,cite,comment,amsfonts,indentfirst,color,setspace,bbold,MnSymbol,arydshln}
\usepackage{mathtext,marvosym,textcomp}
\usepackage[makeroom]{cancel}
\usepackage{mathtools,slashed,xcolor}
\usepackage{blkarray}

\usepackage[debug,pageanchor=false]{hyperref}
\hypersetup{colorlinks=true,linktocpage,breaklinks,
	urlcolor=blue,
	linkcolor=red,
	citecolor=blue
}

\def\a{\alpha} \def\b{\beta} \def\g{\gamma} \def\d{\delta} \def\e{\epsilon}
  \def\h{\eta} 
  \def\k{\kappa} \def\l{\lambda} \def\m{\mu}
\def\n{\nu}  \def\p{\pi}  \def\r{\rho}
 \def\s{\sigma}   \def\f{\varphi}
\def\ff{\phi} \def\c{\chi} \def\y{\psi}

\def\G{\Gamma}

\def\F{\Phi}

\def\fr{\frac}  \def\dt{\partial}

\def\mc{\mathcal}
\def\mH{\mathcal{H}}

\def\tx{\tilde{x}}

\def\tK{\tilde{K}}
\def\te{\tilde{\epsilon}}
\def\tdt{\tilde{\partial}}
\def\mE{\mathcal{E}}
\def\mK{\mathcal{K}}
\def\mF{\mathcal{F}}
\def\mR{\mathcal{R}}

\def\eh{\hat\epsilon}

\def\Tr{\mbox{Tr}}

\def\Tr{\mbox{Tr}}

\def\XX{\mathbb{X}}

\newcommand\bqa {\begin{eqnarray}}
\newcommand\eqa {\end{eqnarray}}

\newcommand{\bear}{\begin{array}}
\newcommand{\enar}{\end{array}}

\def\beq{\begin{equation}}
\def\eeq{\end{equation}}
\def\bea{\begin{eqnarray}}
\def\eea{\end{eqnarray}}

\def\F{{\mathcal{F}}}

\begin{document}

\begin{titlepage}
		
	\phantom{1}

	\vfill

	\begin{center}
		\baselineskip=16pt
		{\Large \bf 
            Fermionic T-duality of DFT
		}
		\vskip 1cm
			Lev Astrakhantsev$^{a,b,c}$\footnote{\tt lev.astrakhantsev@phystech.edu}, Ilya Bakhmatov$^{a,b,d}$\footnote{\tt ibakhmatov@itmp.msu.ru },
		Edvard T. Musaev$^{a}$\footnote{\tt musaev.et@phystech.edu}	
		\vskip .3cm
		\begin{small}
			{\it 
			    $^a$Moscow Institute of Physics and Technology,
		  Russia,\\
				$^b$Institute for Theoretical and Mathematical Physics, Lomonosov Moscow State University, Russia\\
				$^c$Institute of Theoretical and Experimental Physics, 
		  Moscow, Russia\\
				$^d$Kazan Federal University, Institute of Physics, 
		  Kazan,  Russia\\
			}
		\end{small}
	\end{center}
		
	\vfill 
	\begin{center} 
		\textbf{Abstract}
	\end{center} 
	\begin{quote}
         We provide a complete proof that non-abelian fermionic T-duality along a non-anticommuting Killing spinor always generates a solution to double field theory equations. Examples of non-abelian fermionic T-dualities of string backgrounds with non-vanishing b-field are investigated.
	\end{quote} 
	\vfill
	\setcounter{footnote}{0}
\end{titlepage}
	
\clearpage
\setcounter{page}{2}

\section{Introduction}

In its current formulation string theory is not a background independent theory, that is to investigate dynamics of a string one first has to specify the data defining background fields. For consistency of the two-dimensional theory these must satisfy field equations of supergravity. It is now clear that the total space of vacua of the string is degenerate in the sense that there exists a symmetry making different supergravity solutions indistinguishable by the string. Among such duality symmetries the most well known probably is the perturbative T-duality symmetry, which transforms target (hence ``T'' in the name) space-time fields such that the string partition function remains the same \cite{Buscher:1985kb,Buscher:1987qj,Duff:1989tf}. The transformation is performed along a bosonic isometry of the background and at the field theory level is manifested in terms of the so-called Buscher rules \cite{Buscher:1987sk} (see \cite{Giveon:1994fu} for a review). When fermionic isometries are present, i.e.\ a given background is supersymmetric, tree-level string theory can be shown to admit an isometry called fermionic T-duality. This first has been constructed in \cite{Berkovits:2008ic} to provide an interpretation of a duality relating amplitudes of $\mc{N}=4$ SYM and its Wilson loops (see \cite{OColgain:2012si} for a review).  For 2d sigma models with target space-time being a coset space these symmetries have been generalized to a transformation along a superalgebra in \cite{Borsato:2017qsx,Borsato:2018idb} (see also \cite{Bielli:2021hzg,Bielli:2022gmm} for the recent sigma-model results). Bosonic non-abelian T-duality has found numerous applications as a solution generating technique (see \cite{Thompson:2019ipl} for a review).

Recently Buscher rules for non-abelian fermionic T-duality transformation of a general background beyond coset spaces  was formulated in \cite{Astrakhantsev:2021rhj}. The transformation of the $d=10$ type II supergravity fields is superficially the same as in the abelian case~\cite{Berkovits:2008ic}. For the transformed dilaton $\ff$ and the R-R bispinor $\F^{\a\hat\b}$ encoding gauge invariant field strengths we have
\begin{equation}
 \label{fTtrans}
   \F' = \F + 16\,i\, \frac{\e \otimes \hat{\e}}{C},\qquad
   \ff' = \ff + \fr12 \log C,
\end{equation}
where $\e^\a,\hat\e^{\hat\a}$ is a pair of Killing spinors that specify a fermionic isometry direction in $\mathcal{N} = 2$ supersymmetric theory\footnote{For the sake of brevity we only display the expressions for the type IIB case, where both spinors have the same chirality. Everything extends straightforwardly to the case of IIA supergravity. For our spinor and gamma-matrix conventions, see~\cite{Astrakhantsev:2021rhj}.}. What changes in the non-abelian case is the prescription for the parameter~$C$,
\begin{equation}
    \label{eq:naftd}
    \begin{aligned}
        \dt_m C & = i K_m - i b_{mn} \tK^n,\\
        \tdt^m C & = - i \tK^m,
    \end{aligned}
\end{equation}
where
\begin{align}\label{Cdef1}
    K_m &= \e \bar\g_m \e - \eh \bar\g_m \eh,\\
    \label{Cdef2}
    \tK^m &= \e \bar\g_m \e + \eh \bar\g_m \eh.
\end{align}
Note that vanishing of $\tK^m$ is the abelian constraint for the Killing spinors, $[\d_{\e,\eh},\d_{\e,\eh}] = -\tK^m P_m$ by virtue of the supersymmetry algebra. Thus in the non-abelian case $C$ acquires dual coordinate dependence, as well as a contribution from the NSNS 2-form field $b_{mn}$. Usual coordinates $x^m$, that we here refer to as geometric, together with the set of coordinates $\tx_m$, that we here refer to as dual, parameterize the space-time of double field theory. This theory has been first introduced in \cite{Siegel:1993th,Siegel:1993bj} as a proper T-duality covariant description of string theory backgrounds and has been further developed in \cite{Hohm:2010jy,Hohm:2010pp,Hohm:2011dv}. Fields of the theory depend on the full set of coordinates $(x^m,\tx_m)$ given the section condition $\dt_m \bullet \tdt^m \bullet =0$ is satisfied. Bullets denote any field of the theory. Double field theory is covariant under local generalized diffeomorphism transformations, that include standard diffeomorphisms, gauge transformations of the Kalb-Ramond field, T-duality transformation and a set of local transformations, that in particular include $\b$-shifts of \cite{Baron:2022but,Garousi:2022qmk}. For doubled torus the dual coordinates $\tx_m$ have the meaning of the closed string winding modes.

Hence, given the dependence of the function $C$ on both dual and standard coordinates, non-abelian fermionic T-duality in general results in solutions of double field theory. Examples include non-abelian fermionic T-duals of empty Minkowski space-time and of D$p$-brane solutions~\cite{Astrakhantsev:2021rhj}. In general we observe three classes of solutions: i) real backgrounds that depend on dual time; ii) complex backgrounds, that solve supergravity equations; iii) non-geometric complex backgrounds. Note that the non-abelian Killing spinors anticommute to give the vector $\tK^m$, and thus together they form a closed superisometry subalgebra. Hence, the full non-abelian T-duality transformation may be defined, consisting of two steps: the RR-field and the dilaton shift as in \eqref{f1kill} and further (formal) abelian T-duality along $\tK^m$. Due to the actual dependence on the corresponding coordinate, the T-duality transformation is formal in the sense, that this is a reflection in the doubled space of DFT. Under this full T-duality the first class of solutions actually become complex solutions to supergravity equations, as the transformation includes timelike T-duality. Backgrounds of the second class become dependent only on geometric coordinates and again solve supergravity equations. The most tricky are the backgrounds of the third class, that depend on combinations $x\pm \tx$ in such a way that the section constraint is satisfied. This means that there exists a DFT coordinate frame where the combination becomes geometric, however in this frame the background cannot be described in terms of space-time metric and gauge fields. Such backgrounds have been called non-Riemannian in \cite{Jeon:2011cn}. We refer to these as genuinely non-geometric actually adding such backgrounds to those, that cannot be T-dualized to ordinary geometric solutions of supergravity \cite{Dibitetto:2012rk}.

The specific way in which the definition of $C$~\eqref{eq:naftd} was modified by the extra $\tK^m$ terms is based on the analysis of the DFT constraints and equations of motion and is in consistency with the sigma-model analysis for (super)coset target spaces. The dilaton equation was checked in~\cite{Astrakhantsev:2021rhj}, and the prescription was further supported by explicit examples. Here we provide proofs that all DFT equations of motion hold, including the generalized metric and the R-R fields equations.

The organization of this paper is simple: in Section \ref{sec:eom} after a brief review of the DFT formalism we prove that the equations of motion are satisfied after the duality transformation. This is followed by some examples in Section \ref{sec:ex}.

\section{Equations of motion}\label{sec:eom}
 Previously in \cite{Astrakhantsev:2021rhj} it has been shown, that the dilaton equation of motion is invariant under the duality transformation. Here we extend the proof to DFT equations for the generalized metric $\mH_{MN}$, which encodes the NS-NS sector, and for the O(10,10) spinor $|\F\rangle$ representing the R-R sector. In our approach to DFT we follow \cite{Hohm:2010pp,Hohm:2011dv}, where the covariant action for NS-NS fields and the full covariant action including the R-R sector were constructed. Let us briefly go through the relevant conventions. The action of double field theory can be written as
\begin{equation}
\label{act}
S=S_{NSNS}+S_{RR}=\int d^{10}x\, d^{10}\tilde{x} \left( e^{-2d} \mathcal{R} (\mathcal{H},d) + \frac14 (\slashed{\partial}\chi)^{\dagger} S \,\slashed{\partial}\chi \right),
\end{equation}
where the NS-NS degrees of freedom are encoded by the invariant dilaton $d$ and the generalized metric $\mH_{MN}$ with its spin representative $S\in \mathrm{Spin}(10,10)$, while the R-R field strengths are contained in the O(10,10) spinor variable~$\c$. In general the fields are allowed to depend on the full doubled set of coordinates $\XX^M=(x^m,\tx_m)$, assuming that the section constraint is satisfied:
\begin{equation}
    \begin{aligned}
     \h^{MN}\dt_M \bullet \dt_N \bullet = 0, && 
     \h^{MN}=
        \begin{bmatrix}
            0 & \d_m{}^n \\
            \d_n{}^m & 0
        \end{bmatrix}.
    \end{aligned}
\end{equation}
In what follows we will always assume that the non-abelian fermionic T-duality acts on solutions to the ordinary supergravity equations, and hence the initial fields do not depend on the dual coordinates $\tx_m$. Non-abelian fermionic T-dual backgrounds may depend on the dual coordinates, however as we have shown in \cite{Astrakhantsev:2021rhj} the section constraint always holds.

The metric $g_{mn}$ and the Kalb-Ramond field $B_{mn}$ are encoded in the generalized metric $\mH_{MN}$ that is an element of the coset O(10,10)/O(1,9)$\times$O(9,1):
\begin{equation}
 \mathcal{H}_{MN}=\begin{bmatrix} g_{mn}-b_{mp}g^{pq}b_{qn}&b_{mp}g^{pl}\\
-g^{kp}b_{pn}&g^{kl}\end{bmatrix}.
\end{equation}
The dilaton $\phi$ together with $g=\det g_{mn}$ forms the so-called invariant dilaton
\begin{equation}
    d = \phi - \fr14 \log g,
\end{equation}
which transforms as a scalar under DFT symmetries. The DFT curvature scalar $\mathcal{R} (\mathcal{H},d)$ has been first introduced in \cite{Hohm:2011si} as trace of the Ricci curvature $\mathcal{R}_{MN} (\mathcal{H},d)$,
\begin{equation}
\mathcal{R}_{M N} \equiv \frac{1}{4}\left(\delta_{M}^{P}-\mathcal{H}_{M}^{P}\right) \mathcal{K}_{P Q}\left(\delta_{N}^{Q}+\mathcal{H}_{N}^{Q}\right)+\frac{1}{4}\left(\delta_{M}^{P}+\mathcal{H}_{M}^{P}\right) \mathcal{K}_{P Q}\left(\delta_{N}^{Q}-\mathcal{H}_{N}^{Q}\right),
\end{equation}
where 
\begin{equation}
\label{Keq}
    \begin{aligned} 
        \mathcal{K}_{M N} =&\ \frac{1}{8} \partial_{M} \mathcal{H}^{K L}    \partial_{N} \mathcal{H}_{K L}+2 \partial_{M} \partial_{N} d+\big(\partial_{L}-2\left(\partial_{L} d\right)\big)\,\mH^{K L}\,\bigg(\partial_{(M} \mathcal{H}_{N)K} - \fr14\dt_K \mH_{MN} \bigg) \\
        &+\fr14 \Big(\mH^{KL}\mH^{PQ} - 2 \mH^{KQ}\mH^{LP}\Big)\,\dt_K \mH_{MP}\dt_L \mH_{NQ}.
    \end{aligned}
\end{equation}
Note that here we are using the tensor $\mc{K}_{MN}$ in the form of \cite{Rudolph:2016sxe} which gives the same Ricci curvature as that of \cite{Hohm:2011dv}, however proves to be more convenient in explicit calculations. The corresponding Ricci scalar has the form 
\begin{equation}
\label{eq:dilaton}
\begin{aligned}
\mathcal{R}(\mathcal{H},d) &\equiv  4 \mathcal{H}^{M N} \partial_{M} \partial_{N} d-\partial_{M} \partial_{N} \mathcal{H}^{M N}-4 \mathcal{H}^{M N} \partial_{M} d \partial_{N} d+4 \partial_{M} \mathcal{H}^{M N} \partial_{N} d \\
&+\frac{1}{8} \mathcal{H}^{M N} \partial_{M} \mathcal{H}^{K L} \partial_{N} \mathcal{H}_{K L}-\frac{1}{2} \mathcal{H}^{M N} \partial_{M} \mathcal{H}^{K L} \partial_{K} \mathcal{H}_{N L}.
\end{aligned}
\end{equation}

The R-R potentials of the Type II supergravity theories are encoded in the O(10,10) spinor:
\begin{equation}
    |\chi\rangle = \sum_p \fr{1}{p!} C_{m_1\dots m_p}\y^{m_1}\dots \y^{m_p}|0\rangle.
\end{equation}
Here the gamma matrices $(\y^a,\y_a)$ of Spin$(10,10)$ are defined in the usual way (up to rescaling)
\begin{equation}
    \{\y_a,\y^b\}= \d_a{}^b,
\end{equation}
and the Clifford vacuum is defined as usual as $\y_a |0\rangle = 0$. In the covariant spinorial notations the field strengths for the R-R potentials read
\begin{equation}
|F\rangle \equiv |\slashed\partial\chi\rangle=\sum_{p} \frac{1}{p!} F_{m_{1} \ldots m_{p}} \psi^{m_{1}} \cdots \psi^{m_{p}}|0\rangle.
\end{equation}
To define Dirac conjugation one introduces the matrix
$A=\left(\y^{0}-\y_{0}\right)\left(\y^{1}-\y_{1}\right) \cdots\left(\y^{9}-\y_{9}\right)$, that gives
\begin{equation}
\langle F|=\langle{\slashed\partial\chi}|=\sum_p \frac{1}{p !}\langle 0|A \psi^{m_{p}} \cdots \psi^{m_{1}} F_{m_{1} \ldots m_{p}}.
\end{equation}
Finally, the kinetic operator $\mathcal{K}=A^{-1}S$ is written in terms of the Spin(10,10) image of the generalized metric, and it also contributes to the variation with respect to $\mH_{MN}$. 

Here one should be careful regarding which R-R potentials enter the action and hence the equations of motion. Indeed, one distinguishes at least two sets of R-R field strengths: those that transform under T-duality as components of an O(10,10) spinor, denoted $F_{m_1\dots m_p}$ and those that are gauge invariant denoted $\F_{m_1\dots m_p}$. In spinorial notations these two are related as follows
\begin{equation}
    |\F\rangle = e^{\phi}e^{-\fr12 \mathbb{B}}|F\rangle, \quad \mathbb{B} = \fr12b_{mn}\y^{m}\y^{n}.
\end{equation}
Note that the above is not a covariant expression as $b_{mn}$ does not transform as component of a spinor and has rather complicated non-linear transformations involving the space-time metric. We also include the factor $e^{\phi}$ to obtain precisely the field strengths $\mF$, which transform linearly under non-abelian fermionic T-duality. 

Now we are ready to check the invariance of the full set of DFT equations under NAfTD. The equations read
\begin{equation}
\label{equa}
    \begin{gathered}
        \mc{R}  = 0,\\
        \mathcal{R}_{M N}+e^{2d} \mathcal{E}_{MN}=0,\\
        \slashed{\dt}\mathcal{K} \,|\F\rangle=0,
    \end{gathered}
\end{equation}
where $\mathcal{K}$ defines duality relations via $|\F\rangle = - \mathcal{K} \,|\F\rangle$ and the symmetric R-R energy-momentum tensor $\mc{E}_{MN}$ is defined as 
\begin{equation}
    \mE_{MN} = -\fr{1}{16}\mH_{P(M} \langle F|\G_{N)}{}^P |F\rangle. 
\end{equation}
Invariance of the dilaton equation, that is the first line in \eqref{equa} has been shown in \cite{Astrakhantsev:2021rhj}. The R-R fields equation, that is the last line in \eqref{equa}, is simply the integrability condition for the duality relation, given $\slashed{\dt}\slashed{\dt}=0$ on the section condition. Since non-abelian fermionic T-duality does not spoil the duality relations between the R-R $p$-form fields, the same equation as above holds for $\d\F$. Hence, the transformation of the R-R field equations under T-duality vanishes.

To show invariance of the remaining Einstein equation of DFT we must show
\begin{equation}
    \d \mc{R}_{MN} + \d (e^{2d}\mc{E}_{MN})=0,
\end{equation}
where $\d$ denotes the non-abelian fermionic T-duality transformation. The reason why the transformation of $e^{2d}$ must be considered together with the transformation of the energy-momentum tensor will become clear momentarily. Before that, let us write the transformation of the Ricci tensor
\begin{equation}
\label{eq:deltaR}
    \begin{aligned}
        \d \mR_{MN} & = \fr12 \d \mK_{MN} - \fr12 \mH_M{}^P \mH_N{}^Q \d \mK_{PQ}, \\
        \d\mK_{M N} &=\ 2 \partial_{M} \partial_{N} \d d-2(\partial_{L} \d d )\,\mH^{K L}\,\bigg(\partial_{(M} \mathcal{H}_{N)K} - \fr14\dt_K \mH_{MN} \bigg).
    \end{aligned}
\end{equation}
Explicitly substituting fermionic T-duality rules \eqref{eq:naftd}, we obtain the following for the transformation of the components of generalized Ricci tensor:
\begin{equation}
    \begin{aligned}
       C^2\d\mc{R}^{m n}  =&\  - \frac{i}{2}\, C {\nabla}^{(m}{{K}^{n)}}\,   - \frac{1}{2}\, {K}^{m} {K}^{n} + \frac{1}{2}\, {\tK}^{m} {\tK}^{n} ; \\
       C^2\d\mc{R}^{m}{}_{n}   = &\ C^2\d\mc{R}^{m k}b_{nk} + \frac{i}{2}\, C {K}^{k} H_{knl}\,  {g}^{m l}  + i  C  {\nabla}^{m}{{\tK}_{n}}+ {K}^{m} {\tK}_{n}   - {K}_{n} {\tK}^{m}   \\
       &+\frac{i}{2}\, C {\tK}^{k}  H_{klp}\,  {b}_{n q} {g}^{m p} {g}^{l q}  -  i C {\nabla}^{[k}{{K}^{m]}}\,  {b}_{n k};\\
       C^2\d\mc{R}_{m n}   =&\ C^2\d\mc{R}^{kl}\,  {b}_{m k} {b}_{n l}    + \frac{i}{2}\, C {K}^{k} H_{kml}\,  {b}_{n p} {g}^{l p} + i C \nabla^l \tK_m b_{nl} \\
       & + {K}_{m} {\tK}^{k}  {b}_{k n}- {K}^{k} {\tK}_{m}  {b}_{k n} +\fr{i}2\nabla_{(m}K_{n)}- \frac{1}{2}\, {\tK}_{m} {\tK}_{n}  + \frac{1}{2}\, {K}_{m} {K}_{n}, 
    \end{aligned}
\end{equation}
where we have used the Killing vector condition  $\nabla_{(m}\tK_{n)}=0$. In what follows for concreteness we stick to the Type IIB theory, whose supersymmetry transformations read
\begin{equation}
\label{susy-vars}
\begin{aligned}
\d\psi_m &= \nabla_m\e - \frac14 \slashed{H}_m\e - \frac{e^\f}{8} \left( \slashed{F}_{(1)} + \slashed{F}_{(3)} + \frac12 \slashed{F}_{(5)} \right) \bar\g_m \eh \\
\d\hat\psi_m &= \nabla_m\eh + \frac14 \slashed{H}_m \eh + \frac{e^\f}{8} \left( \slashed{F}_{(1)} - \slashed{F}_{(3)} + \frac12 \slashed{F}_{(5)} \right) \bar\g_m \e,\\
\d\l &= \slashed{\dt}\f\,\e -\frac12 \slashed{H} \e + \frac{e^\f}{2} \left( 2\slashed{F}_{(1)} + \slashed{F}_{(3)} \right) \eh,\\
\d\hat\l &= \slashed{\dt}\f\,\eh +\frac12 \slashed{H} \eh - \frac{e^\f}{2} \left( 2\slashed{F}_{(1)} - \slashed{F}_{(3)} \right) \e,
\end{aligned}
\end{equation}
where
\begin{align}
\slashed{F}_{(n)} = \frac{1}{n!} F_{m_1\ldots m_n} \g^{m_1\ldots m_n},\qquad
\slashed{H}_m = \frac12 H_{mnk} \g^{nk}.
\end{align}
Proof for the Type IIA theory goes along precisely the same lines.

Let us now return to the energy-momentum tensor and define
\begin{equation}
    \tilde{\mE}_{MN}= e^{2d}\mE_{MN} = -\fr{1}{16}\fr{1}{\sqrt{g}} \mH_{P(M}\langle \F| e^{-\fr12 \mathbb{B}}\G_{N)}{}^P e^{\fr12 \mathbb{B}}|\F\rangle.
\end{equation}
Now we see, that the factors $e^\phi$ precisely cancel those coming from the transition to gauge invariant field strengths.  The remaining overall factor $\fr{1}{\sqrt{g}}$ will be used to turn the epsilon symbol coming from the Dirac conjugation matrix $A$ into the epsilon tensor.  The only complication here is the exponents of $\mathbb{B}$, which can be rewritten nicely using the Baker-Campbell-Hausdorf formula $e^{-\fr12 \mathbb{B}}\mathcal{X}e^{\fr12 \mathbb{B}}=\mathcal{X}-\fr12[\mathbb{B},\mathcal{X}]+\frac{1}{8}[\mathbb{B},[\mathbb{B},\mathcal{X}]]+\dots$. For that we denote $\tilde{\G}^{MN} = e^{-\fr12 \mathbb{B}}\G^{MN} e^{\fr12 \mathbb{B}}$ and write its components explicitly for further convenience:
\begin{equation}
    \begin{aligned}
        \tilde{\G}^{mn} &= \G^{mn}, \\
        \tilde{\G}^{m}{}_{n} &= \G^{m}{}_{n} - \G^{mk}b_{kn}, \\
        \tilde{\G}_{mn} &= \G_{mn} + 2 \G^k{}_{[m}b_{n]k} + \G^{kl}b_{km}b_{ln}.
    \end{aligned}
\end{equation}
Hence, we have for the tensor $\tilde{\mE}_{MN}$:
\begin{equation}
    \label{eq:tildeE}
    \begin{aligned}
        \tilde{\mE}^{mn} & = -\fr{1}{16} g^{k(m}\langle \F| \G^{n)}{}_{k}|\F\rangle, \\
        \tilde{\mE}^{m}{}_{n} & =  \tilde{\mE}^{mk}b_{n k} -\fr{1}{32}\langle \F| \G_{nk}g^{mk} + \G^{mk}g_{nk}|\F\rangle,\\
        \tilde{\mE}_{mn} & = \tilde{\mE}^{l}{}_{(m}b_{n)l} + \tilde{\mE}^{kl}(b_{mk}b_{nl} - g_{mk}g_{nl}).   
    \end{aligned}
\end{equation}
Now the strategy to prove that the Einstein equation of DFT holds upon a non-abelian fermionic T-duality is to first write the transformation of the generalized Ricci tensor $\d \mR_{MN}$ in the same form as above. What remains then is to simply check the cancelation of $\d \mR^{mn}$ against the first line of \eqref{eq:tildeE} and the corresponding terms of $\d \mR^{m}{}_{n}$ against the last two terms in the second line of \eqref{eq:tildeE}.

To do so, let us first show that the second line in  $C^2\d\mc{R}^{m}{}_{n}$ of \eqref{eq:deltaR} vanishes. For that we use \eqref{susy-vars} to write
\begin{equation}
    \begin{aligned}
        &- i C \nabla^{[k}K^{m]}  = -2 i C \Big(\e \g^{[m}\nabla^{k]}\e - \te \g^{[m}\nabla^{k]}\te\Big) \\
        &= -\fr{i C}{2}\bigg[\e \g^{[m}\slashed{H}{}^{k]}\e + \te \g^{[m}\slashed{H}{}^{k]}\te + \fr12 \e \g^{[m}\Big( \slashed{\F}_1 + \slashed{\F}_3 +\fr12 \slashed{\F}_5\Big)\g^{k]}\te + \fr12 \te \g^{[m}\Big( \slashed{\F}_1 - \slashed{\F}_3 +\fr12 \slashed{\F}_5\Big)\g^{k]}\e\bigg]\\
        &=-\fr{i C}{2}\big(\e \g_q \e + \te \g_q \te\big)H^{kmq} = - \fr{i C}{2} \tK^q H_q{}^{km}=-\frac{i}{2}\, C {\tK}^{q}  H_{qlp}{b}_{n k} {g}^{m p} {g}^{l k},
    \end{aligned}
\end{equation}
where in the second line we used symmetry properties of gamma matrices, $\te \g^{m}\slashed{\F}_{1,5} \g^k \e = \e \g^k \slashed{\F}_{1,5} \g^m \te$ and $\te \g^{m}\slashed{\F}_{3} \g^k \e = -\e \g^k \slashed{\F}_{3} \g^m \te$.  With that we finally obtain the transformation of the generalized Ricci tensor under non-abelian fermionic T-duality:
\begin{equation}
    \begin{aligned}
       C^2\d\mc{R}^{m n} & =  - \frac{i}{2}\, C {\nabla}^{(m}{{K}^{n)}}\,   - \frac{1}{2}\, {K}^{m} {K}^{n} + \frac{1}{2}\, {\tK}^{m} {\tK}^{n} ; \\
       C^2\d\mc{R}^{m}{}_{n} & = C^2\d\mc{R}^{m k}b_{nk}+ \frac{i}{2}\, C {K}^{k} H_{knl}\,  {g}^{m l} + i\, C  {\nabla}^{m}{{\tK}_{n}}+ {K}^{m} {\tK}_{n}   - {K}_{n} {\tK}^{m}   ;\\
      C^2\d\mc{R}_{m n} &  = C^2 \d\mR^{k}{}_{(m}b_{n)k} + C^2 \d\mR^{kl}\big(b_{mk}b_{nl} - g_{mk}g_{nl}\big).
    \end{aligned}
\end{equation}

Comparing this to \eqref{eq:tildeE} we see that to prove preservation of the Einstein equation after non-abelian fermionic T-duality we only have to show that the following is true:
\begin{equation}
    \label{eq:proof1}
    \begin{aligned}
       - \frac{i}{2}\, C {\nabla}^{(m}{{K}^{n)}}\,   - \frac{1}{2}\, {K}^{m} {K}^{n} + \frac{1}{2}\, {\tK}^{m} {\tK}^{n} - \fr{1}{16} g^{k(m}\d \langle \F| \G^{n)}{}_k |\F\rangle& = 0;\\
       \frac{i}{2}\, C {K}^{k} H_{knl}\,  {g}^{m l} +i\, C  {\nabla}^{m}{{\tK}_{n}} + {K}^{m} {\tK}_{n}   - {K}_{n} {\tK}^{m} -\fr{1}{32}\d \langle \F| \G_{nk}g^{mk} + \G^{mk}g_{nk}|\F\rangle &=0.
    \end{aligned}
\end{equation}
The rest will follow as a consequence. For concrete calculations let us proceed with Type IIB theory, for which we  have
\begin{equation}
    \begin{aligned}
        -\fr{1}{4}\fr{1}{\sqrt{g}}g^{k(m}\langle \F | \G^{n)}{}_k|\F\rangle & = \F^{m}\F^{n}+\frac{1}{2}\F^{m}{}_{pq}\F^{n pq}+\frac{1}{2\times4!}\F^{m}{}_{pqrs}\F^{n pqrs}-\frac{1}{2}g^{mn}\sum\limits_{i=1,3}|\F^{(i)}|^2, \\
        -\fr{1}{8}\fr{1}{\sqrt{g}} \langle \F| \G_{nk}g^{mk} + \G^{mk}g_{nk}|\F\rangle & = \F^{m}{}_{n p}\F^{p}+\frac{1}{6}\F^{m}{}_{n pqr}\F^{pqr}.
    \end{aligned}
\end{equation}

To write the transformation of the R-R field strengths in a similar form let us recall trace identities for the gamma matrices relevant for the Type IIB case:
\begin{equation}
    \begin{gathered}
        \fr{1}{16}\Tr[C \g^\m \hat{\g}_\n] = \d_\n{}^\m, \\
        \fr{1}{16}\Tr[C \g^{\m \bar{\n}\r} \g_{\k \bar{\l}\s}] = 3! \d_\k{}^\m{}_\l{}^\n{}_\s{}^\r=-\fr{1}{16}\Tr[C \g^{\bar{\m} \n\bar{\r}} \g_{\k \bar{\l}\s}], \\
        \fr{1}{16}\Tr[C \g^{\bar\m_1 \m_2 \bar\m_3 \m_4 \bar\m_5} \g_{\n_1 \bar\n_2  \n_3 \bar\n_4  \n_5}] = 5! \d_{\n_1 \dots \n_5}^{\m_1 \dots \m_5} + \e^{\m_1\dots \m_5}{}_{\n_1 \dots \n_5} = -\fr{1}{16}\Tr[C \g^{\m_1 \bar\m_2 \m_3 \bar\m_4 \m_5} \g_{\n_1 \bar\n_2  \n_3 \bar\n_4  \n_5}].
    \end{gathered}
\end{equation}
From these we derive
\begin{equation}
    \begin{gathered}
        \fr1{16}\Tr[C \hat{\g}_{\bar\m} \F] = \F_{\m},\\
        \fr1{16}\Tr[C \g_{\bar\m\n\bar{\r}} \F] = -\F_{\m\n\r},\\
      \fr1{16}\Tr[C \g_{\bar\m_1 \m_2 \bar\m_3 \m_4 \bar\m_5} \F] = \F_{\m_1\dots \m_5}.  
    \end{gathered}
\end{equation}

Lets us start with terms linear in $\d \F$  in the first line of \eqref{eq:proof1} for which we have
\begin{equation}
    \begin{aligned}
        -\fr{i}{2}C^{-1} \nabla^{(m}K^{n)} + \fr12\bigg( \d \F^{(m} \F^{n)}   
         +\fr12 \d \F^{(m}{}_{pq } \F^{n)pq} + \fr1{2\times 4!} \d \F^{(m}{}_{pqrs} \F^{n)pqrs} - \fr12 g^{mn} \sum_{p=1,3} \d \F^{i_1\dots i_p}\F_{i_1\dots i_p}\bigg).
    \end{aligned}
\end{equation}
The most convenient way to proceed is by expressing the first term via the R-R field variations using the BPS equations \eqref{susy-vars}:
\begin{equation}
\label{lineqR-R1}
    \begin{aligned}
        &-\fr{i}{2}C^{-1} \nabla^{(m} K^{n)} =  -i  C^{-1} \big(\e \hat\g^{(n} \nabla^{m)} \e - \te \hat \g^{(n} \nabla^{m)} \te \big)\\
        =&\ -\fr{i}{8}  C^{-1} \Big[   \e \hat \g^{(n} \Big(\slashed{\F}_{(1)} + \slashed{\F}_{(3)}+ \fr12 \slashed{\F}_{(5)}\Big)\,\hat\g^{m)} \te +  \te \hat \g^{(n} \Big(\slashed{\F}_{(1)} - \slashed{\F}_{(3)}+ \fr12 \slashed{\F}_{(5)}\Big)\,\hat\g^{m)}\e\Big]\\
        =&\ -\fr{i}{2}   C^{-1}\bigg[ \e \hat \g^{(n} \te \, \F^{m)} + \fr12 \e \hat \g^{pq(n}\hat \e\, \F^{m)}{}_{pq} + \fr{1}{2\times 4!} \e \hat\g^{p_1 \dots p_4 (n}\hat \e \, \F^{m)}{}_{p_1 \dots p_4}- \fr12 g^{mn} \sum_{p=1,3} \e \hat\g^{i_1 \dots i_p}\te\,\F_{i_1\dots i_p} \bigg],
    \end{aligned}
\end{equation}
where in the second line we used properties of the Majorana spinors $\e$ and $\hat{\e}$ and symmetrization in the indices $(mn)$ such as  $\e \g^{(m}\slashed{H}{}^{n)}\e = 0$ to eliminate terms with the NS-NS field strength. The basic properties we use here are
\begin{equation}
    \begin{aligned}
        \e \g^m \te &= \te \g^m \e, \\
        \e \g^{mnk} \te &= -\te \g^{mnk} \e, \\
        \e \g^{m_1\dots m_5} \te &= \te \g^{m_1\dots m_5} \e.
    \end{aligned}
\end{equation}

Now taking into account that $\d\F^{i_{1}\dots i_{p}}= iC^{-1}\e \hat\g^{i_1 \dots i_p}\hat \e$ we finally arrive at
\begin{equation}
    \begin{aligned}
        &-\fr{i}{2}C^{-1} \nabla^{(m} K^{n)} = -\fr12\bigg( \d \F^{(m} \F^{n)}   
         +\fr12 \d \F^{(m}{}_{pq } \F^{n)pq} + \fr1{2\times 4!} \d \F^{(m}{}_{pqrs} \F^{n)pqrs} - \fr12 g^{mn} \sum_{p=1,3} \d \F^{i_1\dots i_p}\F_{i_1\dots i_p}\bigg),
 \end{aligned}
\end{equation}
that precisely cancels against the remaining linear terms in the first line of \eqref{eq:proof1} . 

Next, terms quadratic in the R-R field variation in the first line of \eqref{eq:proof1} must cancel against ${\tK}^{m} {\tK}^{n}-{K}^{m} {K}^{n}$. To see that we write
\begin{equation}
    \begin{aligned}
        & \fr{C^{-2}}{2}(- {K}^{m} {K}^{n} + {\tK}^{m} {\tK}^{n}) = 2 C^{-2} (\e \g^{(m}\e)(\te \g^{n)}\te) = - \fr{2}{16 \times 16} \Tr \big[\d\F \g^{(m}\d\F^T \g^{n)}\big]\\
        &= -\fr{1}{8\times 16} \Tr\bigg[\Big(\d\slashed{\F}_{(1)} + \d\slashed{\F}_{(3)}+ \fr12 \d\slashed{\F}_{(5)}\Big)\,\g^{(m}\,\Big(\d\slashed{\F}_{(1)} - \d\slashed{\F}_{(3)}+ \fr12 \d\slashed{\F}_{(5)}\Big)\g^{n)}\bigg]\\
        &=-\fr14\bigg( \d \F^{(m}\, \d\F^{n)}   
         +\fr12 \d \F^{(m}{}_{pq }\, \d\F^{n)pq} + \fr1{2\times 4!} \d \F^{(m}{}_{pqrs} \,\d\F^{n)pqrs} - \fr12 g^{mn} \sum_{p=1,3} \d \F^{i_1\dots i_p}\,\d\F_{i_1\dots i_p}\bigg),
    \end{aligned}
\end{equation}
where in the second line we used Fierz identities and the fact that $\g^m,\g^{mnkpl}$ are symmetric and $\g^{mnk}$ is antisymmetric.

Next, check the linear part in the second line of \eqref{eq:proof1}. Similarly to \eqref{lineqR-R1}, we express the covariant derivative in terms of the BPS equations \eqref{susy-vars}:
\begin{equation}
    \begin{aligned}
        & i C \nabla^{m}\tK_{n}  = 2g_{np} i C \Big(\e \g^{p}\nabla^{m}\e + \te \g^{p}\nabla^{m}\te\Big) \\
        &= \fr{i C}{2}g_{np}\bigg[\e \g^{p}\slashed{H}{}^{m}\e - \te \g^{p}\slashed{H}{}^{m}\te + \fr12 \e \g^{p}\Big( \slashed{\F}_1 + \slashed{\F}_3 +\fr12 \slashed{\F}_5\Big)\g^{m}\te - \fr12 \te \g^{p}\Big( \slashed{\F}_1 - \slashed{\F}_3 +\fr12 \slashed{\F}_5\Big)\g^{m}\e\bigg]\\
        &= \fr{i C}{2} H^{m}{}_{nq}K^q-\fr{i C}{4}\Big( \e \hat \g^{p} \te \, \F^{m}{}_{np} +\e \hat \g^{m}{}_{np}\hat \e\, \F^{p} + \fr{1}{6} \e \hat\g^{pqr}\hat \e \, \F^{m}{}_{npqr}+\fr{1}{6}\e \hat \g^{m}{}_{npqr}\hat \e\, \F^{pqr}\Big).
    \end{aligned}
\end{equation}
Taking into account that $\d\F^{i_{1}\dots i_{p}}= iC^{-1}\e \hat\g^{i_1 \dots i_p}\hat \e$ we see these precisely are the remaining linear terms, but with the opposite sign.

The remaining cancellation in \eqref{eq:proof1} to be proven involves the quadratic terms in the second line:
\begin{equation}
    \begin{aligned}
        & \fr{C^{-2}}{2}({K}^{m} \tK_{n} - K_{n} {\tK}^{m}) = 2 C^{-2}g_{np} (\e \g^{[m}\e)(\te \g^{p]}\te) = - \fr{2g_{np}}{16 \times 16}\, \Tr \big[\d\F\, \g^{[m}\d\F^T \g^{p]}\big]\\
        &= -\fr{g_{np}}{8\times 16}\, \Tr\bigg[\Big(\d\slashed{\F}_{(1)} + \d\slashed{\F}_{(3)}+ \fr12 \d\slashed{\F}_{(5)}\Big)\,\g^{[m}\,\Big(\d\slashed{\F}_{(1)} - \d\slashed{\F}_{(3)}+ \fr12 \d\slashed{\F}_{(5)}\Big)\,\g^{p]}\bigg]\\
        &=-\fr14\bigg( \d\F^{m}{}_{n p}\,\d\F^{p}+\frac{1}{6}\,\d\F^{m}{}_{n pqr}\,\d\F^{pqr}\bigg).
    \end{aligned}
\end{equation}
Hence we have shown that the DFT equation of motion for the generalized metric holds true after non-abelian fermionic T-duality defined as \eqref{fTtrans}.

This concludes our proof that non-abelian fermionic T-duality preserves the DFT equations of motion. So a generic supergravity solution becomes a solution of Double Field Theory after the non-abelian fermionic T-duality.

\section{Examples with non-vanishing b-field}\label{sec:ex}

Some examples of transformed solutions have been considered in~\cite{Astrakhantsev:2021rhj}, all of them with vanishing b-field. It has been observed that non-abelian fermionic T-duals of the D$p$-brane solutions are characterized by the function $C$ that depends on coordinates along the brane world-volume only. This is a rather non-trivial observation, since the Killing spinors to the contrary only depend on the harmonic function of the transverse distance from the source brane. However, this dependence is always canceled out by similar contributions coming from the vielbein and its inverse, leaving only linear dependence on the world-volume coordinates and their duals. Let us now inspect what happens in the case of backgrounds with no RR fields but non-vanishing b-field. As an example of such a background we proceed with the type II fundamental string, given by
\begin{equation}
    \begin{aligned}
        ds^2 & = H^{-1}(-dt^2 + dy^2) + dx_{(8)}^2,\\
        b_{ty} & = H^{-1} - 1, \quad e^{-2\ff}   = H e^{-2\ff_0}, \\
        H& = 1+ \fr{h}{|x_{(8)}|^6}.
    \end{aligned}
\end{equation}
This is a 1/2-BPS solution and hence preserves half of the total Type II supersymmetry. The corresponding Killing spinors for type IIA and IIB can be written collectively as
\begin{equation}
    \begin{aligned}
    \label{f1kill}
        \begin{pmatrix}\e \\ \hat{\e}\end{pmatrix} &= H^{-\fr14} \begin{pmatrix}\e_0 \\ \hat{\e}_0\end{pmatrix}, \quad (1 + \G^{01}\mc{O})\begin{pmatrix}\e_0 \\ \hat{\e}_0\end{pmatrix} = 0,\\
        \mc{O} &= \left\{ 
            \begin{aligned}
                \G_{11}, \quad \mathrm{IIA}, \\
                \s^3, \quad \mathrm{IIB}.
            \end{aligned}
        \right.
    \end{aligned}
\end{equation}
Here $\G_{11}$ is the 32$\times$32 gamma matrix in 10 dimensions, while the  Pauli matrix $\s^3$ acts on the column of two Type IIB spinors of the same chirality. Explicitly for the solution as written above we have 
\begin{equation}
    \begin{aligned}
        \e_0 & = (1-\g^{0\bar{1}}) \h=(1-\g^{0}\g^{1})\h \quad \mathrm{IIA,IIB};\\
     \hat{\e}_0 & = \left\{\,
     \begin{aligned}
     &(1+\g^{\bar{0}1})\bar{\h}=(1-\g^{0}\g^{1})\bar{\h}\quad \mathrm{IIA},\\
     &(1+\g^{0\bar{1}}) \h=(1+\g^{0}\g^{1})\h \quad \mathrm{IIB},
     \end{aligned}
     \right.\\
    \end{aligned}
\end{equation}
for an arbitrary constant 16-component spinors $(\h,\bar{\h})$. Notice $\g^{0\bar{1}}=-\g^{\bar{0}1}=\g^{0}\g^{1}$. 

Let us now enumerate all 32 spinors as in \cite{Astrakhantsev:2021rhj} introducing a basis $\{\e_i,\eh_{i}\}$ of 16-component spinors of the opposite chiralities. Then we may write Killing spinors $\e_0$, $\eh_0$ of the Type IIA string in the following most general form
\begin{equation}
    \begin{aligned}
        \e_0 & = \fr14 e^{-i \fr \p 4}H^{-\fr14}(1 - \g_{01}) \sum_{i=1}^{8}a_i \e_i, \\
        \eh_0 & = \fr14 e^{i \fr \p 4}H^{-\fr14}(1 - \g_{01}) \sum_{i=1}^{8}b_i \eh_{i+8},
    \end{aligned}
\end{equation}
where the overall numerical prefactors have been chosen for further convenience and coefficients $a_i$ and $b_i$ are constant.
This yields the following function $C$
\begin{equation}
\label{defCfun}
    C= \fr12(A+B)\big(x^{1}+\tilde{x}_{0}\big)+\fr12(A-B)\big(x^{0}-\tilde{x}_{1}\big),
\end{equation}
where we define
\begin{equation}
    \begin{aligned}
    4A&=\sum_{i=1}^{8}a_{i}^2,\\
    4B&=\sum_{i=1}^{8}b_{i}^2.
    \end{aligned}
\end{equation}
As expected, the section condition of DFT is satisfied even though there seemingly is a dependence on both a coordinate and on its dual. As we have discussed previously in \cite{Astrakhantsev:2021rhj} this simply means, that one can choose such a coordinate frame in the doubled space, where $C$ depends only on a (new) coordinate and not on its dual. Note however, that this does not guarantee that the background is geometric as there might not be possible to define a space-time metric in such frame. This has been precisely the case for one of the examples of \cite{Astrakhantsev:2021rhj}.

Hence, one observes here several distinct possibilities: i) a background that depends on the dual time $\tx_0$, which appears to be real; ii) a background with no dependence on the dual time, which is always complex; iii) a non-geometric background with dependence on a pair $x+\tx$, which can be rotated into a frame where no space-time metric can be defined. For concreteness, let us provide explicit examples of all these possibilities. Conceptually, they repeat those already considered in \cite{Astrakhantsev:2021rhj} and are here mainly to illustrate that the general principle does not change for backgrounds with non-vanishing b-field.  

\textbf{Real background example}

To obtain a formally real background we set  $a_{1}=b_{1}=2$, $A=B=1$ that gives
\begin{equation}
 C=x^{1}+\tilde{x}_{0}.
\end{equation}
In this case for the non-abelian fermionic T-dual background we obtain 
\begin{equation}
    \begin{aligned}
        e^{-2\ff}&=\frac{He^{-2\ff_{0}}}{x^{1}+\tilde{x}_{0}},\\
        m&=\frac{e^{-\ff_{0}}}{2(x^{1}+\tilde{x}_{0})^{3/2}},\\
        \F_{(2)}&=\frac{e^{-\ff_{0}}}{2HC^{3/2}}dx^{01},\\
        \F_{(4)}&=\frac{e^{-\ff_{0}}}{C^{3/2}}\Big((dx^{34}-dx^{27}-dx^{89})\wedge(dx^{56}-dx^{89})+(dx^{23}-dx^{47})\wedge(dx^{58}+dx^{69}-dx^{47})\\
        &+(dx^{24}+dx^{37})\wedge(dx^{59}-dx^{68})\Big),   
    \end{aligned}
\end{equation}
where we explicitly write only the RR field strength and the dilaton as the remaining fields stay the same. We use the obvious notation $dx^{ij} = dx^i\wedge dx^j$. Notice that the 0-form field strength $\F_0$, normally called the Roman's mass $m$, acquires a dependence on a world-volume and a dual coordinate. 

\textbf{Real background example with zero mass}

Non-vanishing Roman's mass in the previous example implies that upon timelike T-duality the background is a solution of the deformed, massive, Type IIA theory. To arrive at solutions to the massless Type II supergravity we choose $a_{1}=b_{2}=2$, $A=B=1$, that gives:
\begin{equation}
    \begin{aligned}
    e^{-2\ff}&=\frac{He^{-2\ff_{0}}}{x^{1}+\tilde{x}_{0}},\\
    \F_{(2)}&=-\frac{e^{-\ff_{0}}}{2C^{3/2}}\Big[dx^{67}+dx^{38}+dx^{49}-dx^{25} \Big],\\
    \F_{(4)}&=\frac{e^{-\ff_{0}}}{2C^{3/2}}\Big[\frac{1}{H}dx^{01}\wedge(dx^{67}-dx^{25}+dx^{38}+dx^{49})+(dx^{89}-dx^{34})\wedge(dx^{26}+dx^{57})\\
    &+(dx^{39}-dx^{48})\wedge(dx^{27}-dx^{56}) \Big].
    \end{aligned}
\end{equation}

\textbf{Complex background example}

To avoid dependence on the dual time $\tx_0$ one may set e.g.  $a_{1}=-ib_{1}=2$, $A=-B=1$ that gives $C=x^{0}-\tilde{x}_{1}$ and  results in the following complex background
\begin{equation}
    \begin{aligned}
        e^{-2\ff}&=\frac{He^{-2\ff_{0}}}{x^{0}-\tilde{x}_{1}}, \\
        m&=\frac{ie^{-\ff_{0}}}{2(x^{0}-\tilde{x}_{1})^{\frac{3}{2}}},\\
        \F_{(2)}&=\frac{ie^{-\ff_{0}}}{2HC^{3/2}}dx^{01},\\
        F_{(4)}&=\frac{ie^{-\ff_{0}}}{C^{3/2}}\Big((dx^{34}-dx^{27}-dx^{89})\wedge(dx^{56}-dx^{89})+(dx^{23}-dx^{47})\wedge(dx^{58}+dx^{69}-dx^{47}) \\
        &+(dx^{24}+dx^{37})\wedge(dx^{59}-dx^{68})\Big).
    \end{aligned}
\end{equation}
Again the Roman's mass is non-vanishing and depends on coordinates $x^0$ and $\tx_1$. As before, with a different choice of the parameters $a_i$ and $b_i$ one can end up with a massless background.

\textbf{Non-geometric example}

Finally, to arrive at a non-geometric background it is enough to set either $A$ or $B$ to vanish, or more generally keep $|A|\neq|B|$. An example is provided by the function
$$
C=x^{1}+x^{0}+\tilde{x}_{0}-\tilde{x}_{1},
$$
with the corresponding Killing spinor choice yielding vanishing RR fields. Same as in \cite{Astrakhantsev:2021rhj}, by non-geometric we mean DFT solutions which cannot be bosonic T-dualized into any geometric background. Following the nomenclature of~\cite{Dibitetto:2012rk} these belong to the so-called genuinely non-geometric class, to which we also include DFT backgrounds, that either depend on dual coordinate or become non-Riemannian in the sense of \cite{Jeon:2011cn}.

Non-abelian fermionic T-dualization of the Type IIB fundamental string background does not add new information to the Type IIA examples considered above. The same is true for more simple backgrounds such as e.g.\ Minkowski space with Kalb-Ramond field given by a pure gauge term.

\section{Summary}

In this work we complete the proof that non-abelian fermionic T-duality generates backgrounds that are always a solution to equations of double field theory. As it has been discussed previously in \cite{Astrakhantsev:2021rhj} non-abelian T-duality along a superisometry consisting of a Killing spinor and a Killing vector can be understood as a two-step process: i)  the purely fermionic transformation according to the rule \eqref{fTtrans}, ii) abelian bosonic T-duality along the Killing vector. Although this is a symmetry of the 2d sigma-model (see e.g. \cite{Borsato:2018idb}) explicit examples show that in general one does not end up with a supergravity background, however, the result is always a solution to DFT equations. 

Non-abelian T-duality along bosonic isometries can also be decomposed in a similar fashion: i) a shift of the b-field linear in dual coordinates, ii) formal abelian T-dualities along all direction \cite{Musaev:2020bwm}. As in the present case, the first step alone produces solutions to double field theory equations that depend on dual coordinates. However, the whole bosonic NATD procedure always ends up with a (generalized) supergravity solution, as all dual coordinates get dualized. In contrast, in the fermionic case one finds examples that depend on combinations $x\pm \tx$ of a geometric coordinate $x$ and its dual $\tx$. Given the section condition is satisfied, there exists such DFT coordinate frame, where the involved combination can be understood as a new geometric coordinate, say $x'=x+\tx$, the dual then would be $\tx'=x-\tx$. However, in this case the space-time metric $g_{mn}$ often cannot be recovered, as the corresponding block of the generalized metric is degenerate, and hence the background is non-Riemannian in the sense of \cite{Jeon:2011cn}. The lack of description of such backgrounds in terms of a space-time metric settles it outside of the set of supergravity solutions. To our knowledge, no examples of a similar behavior are known in the bosonic case, which would be interesting to search for.

Our discussion both here and in \cite{Astrakhantsev:2021rhj} has been restricted by simple isometry superalgebras containing a single fermionic generator and a single bosonic generator. The word ``non-abelian'' therefore refers to the property of the Killing spinor, which does not commute with itself as it is normally required for fermionic T-duality. As we already show in \cite{Astrakhantsev:2021rhj} such transformation can never generate a real background, that is either one ends up with complex valued fields, or with real fields and dependence on dual time. In the latter case further timelike T-duality turns the fields complex. It is tempting to consider more general setups, where the bosonic isometry subgroup is non-abelian and contains more generators. That should provide enough freedom to generate real backgrounds by fermionic dualities. For alternative although similar approach see \cite{Godazgar:2010ph}. Also such more complicated dualization schemes might be useful for searches of self-duality of AdS${}_4\times \mathbb{CP}^3$ \cite{Colgain:2016gdj}.

\section*{Acknowledgments}
This work has been supported by Russian Science Foundation under the grant RSCF-20-72-10144.

\bibliographystyle{utphys}
\bibliography{bib}

\end{document}